\documentclass[10pt]{article}
\usepackage{graphicx}

\def\Title#1{\begin{center} {\Large #1 } \end{center}}
\def\Author#1{\begin{center}{ \sc #1} \end{center}}
\def\Address#1{\begin{center}{ \it #1} \end{center}}

\newcommand\pubblock{\rightline{\begin{tabular}{l} Proceedings of the Fifth Annual LHCP\\ \pubnumber\\
         \pubdate  \end{tabular}}}

\newenvironment{Abstract}{\begin{quotation} \begin{center} 
             \large ABSTRACT \end{center}\bigskip 
      \begin{center}\begin{large}}{\end{large}\end{center} \end{quotation}}

\newenvironment{Presented}{\begin{quotation} \begin{center} 
             PRESENTED AT\end{center}\bigskip 
      \begin{center}\begin{large}}{\end{large}\end{center} \end{quotation}}

\def\beq{\begin{equation}}
\def\eeq#1{\label{#1}\end{equation}}
\def\eeqn{\end{equation}}

\def\beqa{\begin{eqnarray}}
\def\eeqa#1{\label{#1}\end{eqnarray}}
\def\eeqan{\end{eqnarray}}

\let\bar=\overbar

\def\Dslash{\not{\hbox{\kern-4pt $D$}}}
\def\dslash{\not{\hbox{\kern-2pt $\del$}}}

\def\mt{m_t}

\def\msb{{\bar{\ssstyle M \kern -1pt S}}}

\def\pt{\ensuremath{p_{\mathrm{T}}}}
\def\et{\ensuremath{E_{\mathrm{T}}}}
\def\mt{\ensuremath{m_{\mathrm{T}}}}

\usepackage{color}
\usepackage{amsmath}
\usepackage{subfig}
\usepackage{graphicx}

\newcommand\pubnumber{ ATL-PHYS-PROC-2017-095 }

\newcommand\pubdate{\today}

\def\affiliation{
on behalf of the ATLAS Collaboration, \\
Kirchhoff-Institute for Physics, \\
Heidelberg University, Germany}

\begin{document}
\large
\begin{titlepage}
\pubblock

\vfill
\Title{Studying $WW\gamma$ and $WZ\gamma$ production in proton--proton collisions at $\sqrt{s} = 8$~TeV with the ATLAS experiment}
\vfill

\Author{ Julia Isabell Djuvsland }
\Address{\affiliation}
\vfill
\begin{Abstract}
Quartic gauge couplings are tested by this study of the production of $WW\gamma$ and $WZ\gamma$ events in 20.2~fb$^{-1}$ of 
proton--proton collisions at a centre-of-mass energy of $\sqrt{s} = 8$~TeV recorded with the ATLAS detector at the LHC. 
The final state of $WW\gamma$ events containing an electron, a muon and a photon is analysed as well as the final states of 
$WW\gamma$ and $WZ\gamma$ production containing an electron or a muon, two jets and a photon. 
For all final states two different fiducial regions are defined: one yielding the best sensitivity to the production 
cross-section of the process and one optimised for the detection of new physical phenomena. 
In the former region, the $WW\gamma$ production cross-section is computed and in both regions, upper limits on the $WW\gamma$ and 
$WZ\gamma$ production cross-section are derived.
The results obtained in the second phase space are combined for the interpretation in the context of anomalous quartic gauge 
couplings using an effective field theory.
\end{Abstract}
\vfill

\begin{Presented}
The Fifth Annual Conference\\
 on Large Hadron Collider Physics \\
Shanghai Jiao Tong University, Shanghai, China\\ 
May 15-20, 2017
\end{Presented}
\vfill
\end{titlepage}
\def\thefootnote{\fnsymbol{footnote}}
\setcounter{footnote}{0}
%

\normalsize 


\section{Introduction}

\begin{figure}[t]
 \centering \quad
 \subfloat[]{\includegraphics[width=.3\textwidth]{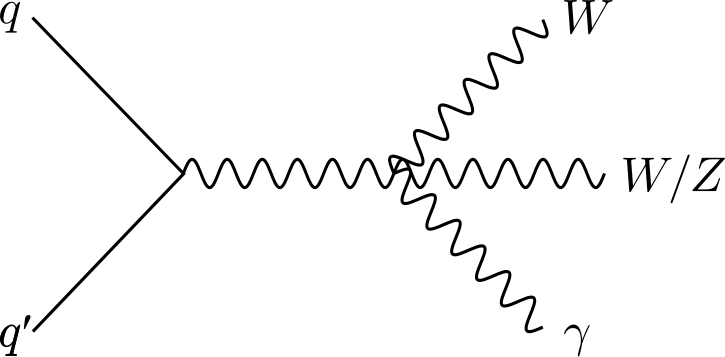}\label{fig:quartFeynm}}\quad
 \subfloat[]{\includegraphics[width=.3\textwidth]{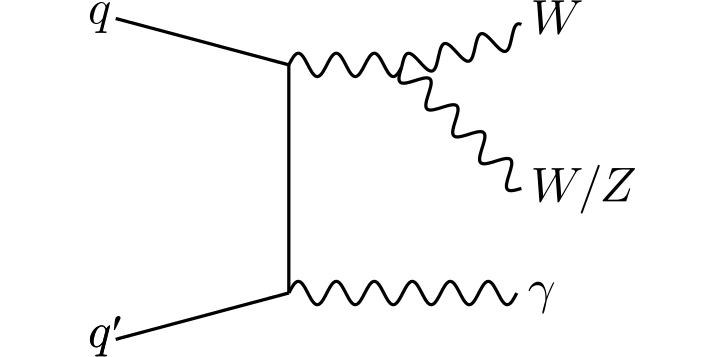}\label{fig:triplFeynm}}\quad
 \subfloat[]{\includegraphics[width=.3\textwidth]{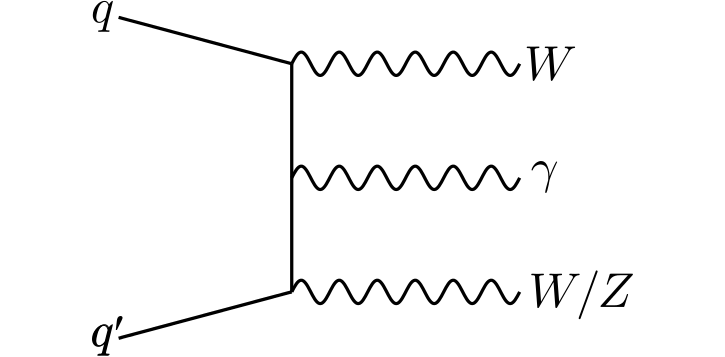}\label{fig:radiFeynm}}\quad
 \caption{Examples of Feynman diagrams showing $WW\gamma$ and $WZ\gamma$ production. The quartic vertex is shown in 
\protect\subref{fig:quartFeynm}, while the production including radiative processes is depicted 
in \protect\subref{fig:triplFeynm} and \protect\subref{fig:radiFeynm}. From Reference~\cite{Aaboud:2017tcq}.}
  \label{fig:feynmans}
\end{figure}

The strong success of the Standard Model (SM) of particle physics to explain all elementary particles and their interactions is 
slightly weakened by its shortcomings. 
Therefore, the model is tested with scrutiny to find possible extensions needed to describe particle interactions at high 
energies. 
An interesting test is the study of proton--proton collisions that produce three electroweak gauge bosons. 
The couplings of these processes are completely determined by the SM, such that any observed deviation form the expectations 
would directly hint to new physical phenomena. 
The analysis of triboson production tests in particular the non-Abelian gauge structure of the SM, as the production of three 
gauge bosons includes a quartic vertex as depicted for example in Figure~\ref{fig:quartFeynm}.
Five different triboson combinations have been studied at the LHC so far: $W\gamma\gamma$~\cite{Aad:2015uqa, 
Sirunyan:2017lvq}, $Z\gamma\gamma$~\cite{Aad:2016sau, Sirunyan:2017lvq}, $WWW$~\cite{Aaboud:2016ftt} and $WW\gamma$ together with 
$WZ\gamma$~\cite{Chatrchyan:2014bza}. 
The study presented here analyses $WW\gamma$ and $WZ\gamma$ production using the full proton--proton data set recorded with the 
ATLAS detector~\cite{Aad:2008zzm} at a centre-of-mass energy of $\sqrt{s} = 8$~TeV corresponding to an integrated luminosity of 
20.2~fb$^{-1}$. 
This study extends the current results by also analysing the leptonic decay mode of both heavy gauge boson ($e\nu\mu\nu\gamma$ 
final state), while the analysis presented in Reference~\cite{Chatrchyan:2014bza} only exploits final states containing a single 
lepton ($e\nu jj\gamma$ and $\mu\nu jj\gamma$ final states, where $j$ indicates a jet). 
The analysis of the $e\nu\mu\nu\gamma$ final state is only sensitive to $WW\gamma$ production.
The analysis of $e\nu jj\gamma$ and $\mu\nu jj\gamma$ final states is sensitive to $WW\gamma$ and $WZ\gamma$ production but 
cannot distinguish them due to the limited jet energy resolution of the ATLAS detector. 
In the following, an overview of the analysis and the results is given, details can be found in Reference~\cite{Aaboud:2017tcq}.

\section{Signal and Backgrounds}
The iconic feature of this analysis is that it is sensitive to the quartic gauge coupling vertex. Yet, the detector signature for 
$WW\gamma$ and $WZ\gamma$ triboson production including radiative processes -- shown in Figures~\ref{fig:triplFeynm} 
and~\ref{fig:radiFeynm} -- is the same, such that all production modes are considered as signal.
For each final state a signal region is defined by selection criteria that are optimised to yield the best signal significance. 
They are listed in Table~\ref{tab:extFidReg}. 
The events must contain the defining particles, i.e. leptons and a photon, of the respective final state, where the requirement on 
the transverse momenta (\pt) of the leptons arises from the constraints imposed for triggering the events. 
Requiring a sizeable amount of transverse energy (\et) of the photons discards mainly events containing photons from initial or 
final state radiation that do not involve the interesting quartic gauge coupling vertex. 
The selection criteria of the $e\nu\mu\nu\gamma$ final state do not allow reconstructed jets in the events, while in the other 
two channels at least two energetic jets need to be present due to the hadronic decay of one of the heavy gauge bosons. No jets 
originating from $b$-quarks are allowed in the final states. 
Since the leptonic decay of the gauge bosons includes a neutrino that cannot be detected, the signal events have a momentum 
imbalance in the plane transverse to the beam axis ($E_{\mathrm{T}}^{\mathrm{miss}}$). 
It is computed from all calibrated objects in the events and additional energy not attributed to an object is taken into 
account by adding the unassociated tracks. 
The analysis of the $e\nu\mu\nu\gamma$ channel employs a slightly different variable ($E_{\mathrm{T,\,rel}}^{\mathrm{miss}}$) that 
also takes the angular separation of the object closest to 
$E_{\mathrm{T}}^{\mathrm{miss}}$ into account in order to mitigate mismeasurement.
The selection criteria of the $e\nu\mu\nu\gamma$ final state also include a minimum requirement on the invariant dilepton mass  
$m_{e\mu}$ in order to suppress Drell-Yan processes. 
The selection criteria of the $e\nu jj\gamma$ and $\mu\nu jj\gamma$ final states also impose a threshold on the transverse mass 
(\mt) and restrict the invariant mass of the two leading jets in the event ($m_{jj}$) to discard events where these jets do not 
originate from the decay of a heavy gauge boson. 

\begin{table}[t]
\begin{center}
\begin{tabular}{ll}
    $e\nu\mu\nu\gamma$                                           & $e\nu jj\gamma$ or $\mu\nu jj\gamma$         \\    
    \hline          
    1 electron and 1 muon ($\pt > 20$\,GeV) with opposite charge & 1 electron or 1 muon ($\pt > 25$\,GeV)       \\
    no $3^{\mathrm{rd}}$ lepton ($\pt > 7$\,GeV)                 & no $2^{\mathrm{nd}}$ lepton ($\pt > 7$\,GeV) \\
    $\geq 1$ isolated photon ($\et > 15$\,GeV)                   & $\geq 1$ isolated photon ($\et > 15$\,GeV)   \\      
    no jets ($\pt > 25$\,GeV)                                    & $\geq 2$ jets ($\pt > 25$\,GeV), no $b$-jets \\
    $E_{\mathrm{T,\,rel}}^{\mathrm{miss}} > 15$\,GeV             & $E_{\mathrm{T}}^{\mathrm{miss}} > 30$\,GeV   \\
    $m_{e\mu}> 50$\,GeV                                          & $\mt >  30$\,GeV                             \\
                                                                 & 70\,GeV $< m_{jj} <$ 100\,GeV                \\
    \hline
\end{tabular} 
\caption{Selection criteria of the $e\nu\mu\nu\gamma$, $e\nu jj\gamma$ and $\mu\nu jj\gamma$ signal regions.}
\label{tab:extFidReg}
\end{center}
\end{table}

With these selection criteria, a signal purity of 45\% is expected for the analysis of the $e\nu\mu\nu\gamma$ final state. 
The main background processes are the production of top quark pairs in association with a photon ($t\bar{t}\gamma$) and Drell-Yan 
processes with photon radiation ($Z\gamma$). 
In addition, processes with a misidentified object need to be considered, as the signal cross-section is rather low. 
Thus, $WZ$ production can mimic the signal in case an electron is misidentified as a photon (fake $\gamma$ from $e$) or $WW$ and 
$t\bar{t}$ production can be mistakenly identified as signal in case a jet is misidentified as a photon (fake $\gamma$ from jets).
For the analysis of the $e\nu jj\gamma$ ($\mu\nu jj\gamma$) final state, the expected signal purity reduces to 2.5\% (2.8\%) due 
to the large background from $W\gamma$ production in association with jets ($W\gamma + \mathrm{jets}$). 
In addition, the fake $\gamma$ from jets and the fake $\ell$ from jets backgrounds have a sizeable contribution as does 
$t\bar{t}\gamma$ production. 

In this study, the backgrounds are estimated using a mix of data based methods and Monte Carlo simulations. 
For the analysis of the $e\nu\mu\nu\gamma$ final state, the estimation of the fake $\gamma$ from $e$, the fake $\gamma$ from jets 
and the fake $e$ from jets background contributions employs collision data. 
For the estimation of the former, the rate of misidentifying an electron as a photon is measured in $Z \rightarrow ee$ decays and 
the rate in Monte Carlo simulation is corrected to this value.
The other two fake object categories are estimated simultaneously by combining two 2D side band methods. 
Six background enriched control regions are defined using the isolation energy of the electrons and photons as well as the 
identification criteria of the photons and criteria selecting events that contain electrons and jets. 
The novel approach of this method is that the observed number of events in these six control regions and the signal region are 
combined using a likelihood formalism for the dependency of the event counts in the different regions.

For the analysis of the $e\nu jj\gamma$ and $\mu\nu jj\gamma$ final states, the estimation of the fake $\gamma$ from $e$, the 
fake $\gamma$ from jets, the fake $\ell$ from jets and the $W\gamma + \mathrm{jets}$ backgrounds employs collision data. 
The former is estimated in the same way as in the $e\nu\mu\nu\gamma$ channel and the latter three components are again estimated 
simultaneously. 
This way, the interdependence of the different background estimation methods is overcome. 
To this end, three different fits are combined, namely $i$) a binned maximum likelihood fit of the $m_{jj}$ distribution, $ii$) a 
binned maximum likelihood fit of the $E_{\mathrm{T}}^{\mathrm{miss}}$ distribution, and $iii$) a 2D side band method. 
All phase spaces employed in the fit have the $m_{jj}$ requirement inverted with respect to the signal region. 
The background contributions are estimated in these phase spaces and their yield in the signal region is extrapolated using 
the $m_{jj}$ distributions of the different background components.

\begin{figure}[htbp]
 \centering
 \subfloat[$e\nu\mu\nu\gamma$ final state]{\includegraphics[width=.333\textwidth]{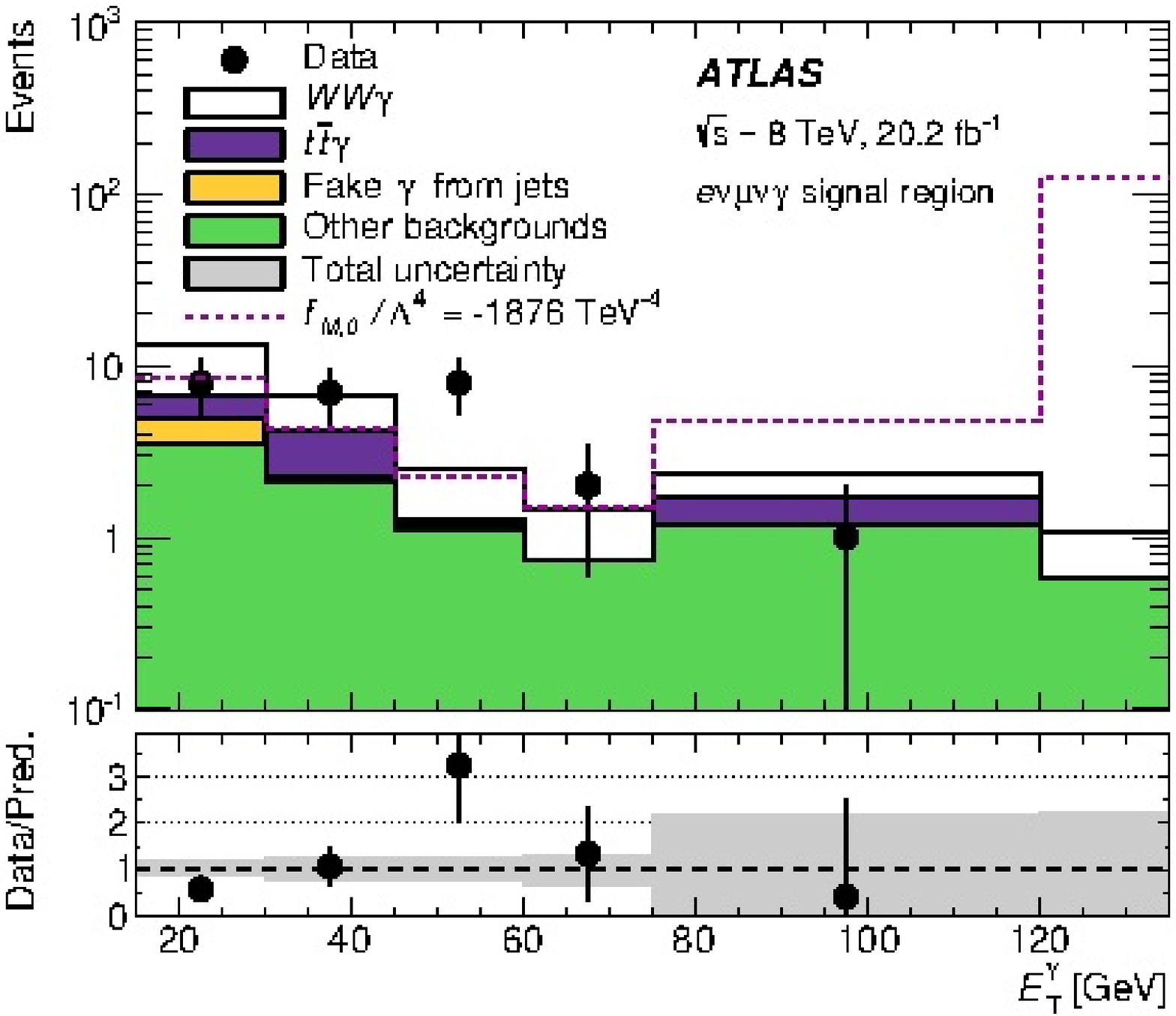}\label{fig:emy}} 
 \subfloat[$e\nu jj\gamma$ final state]{\includegraphics[width=.333\textwidth]{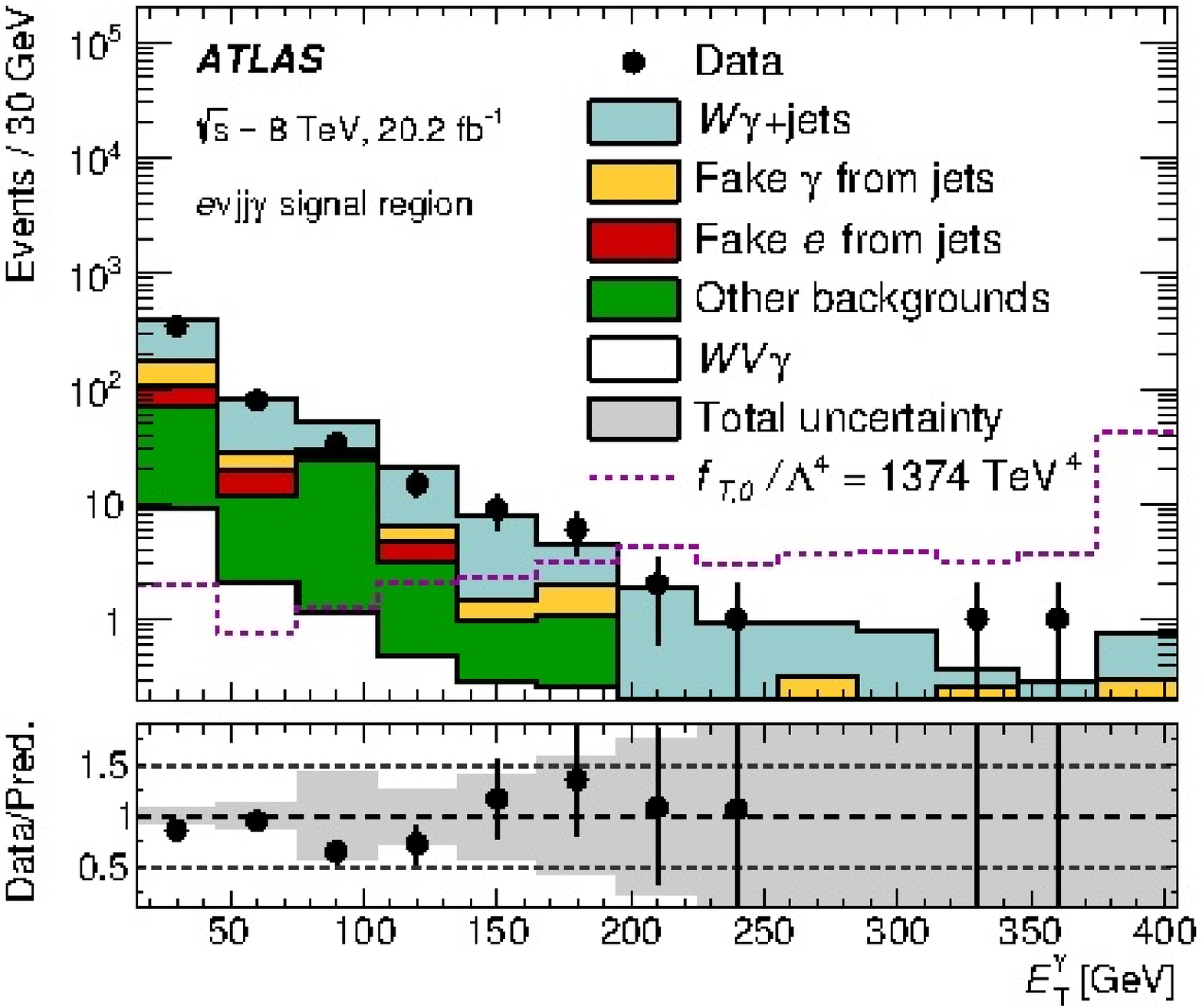}\label{fig:ejjy}}
 \subfloat[$\mu\nu jj\gamma$ final state]{\includegraphics[width=.333\textwidth]{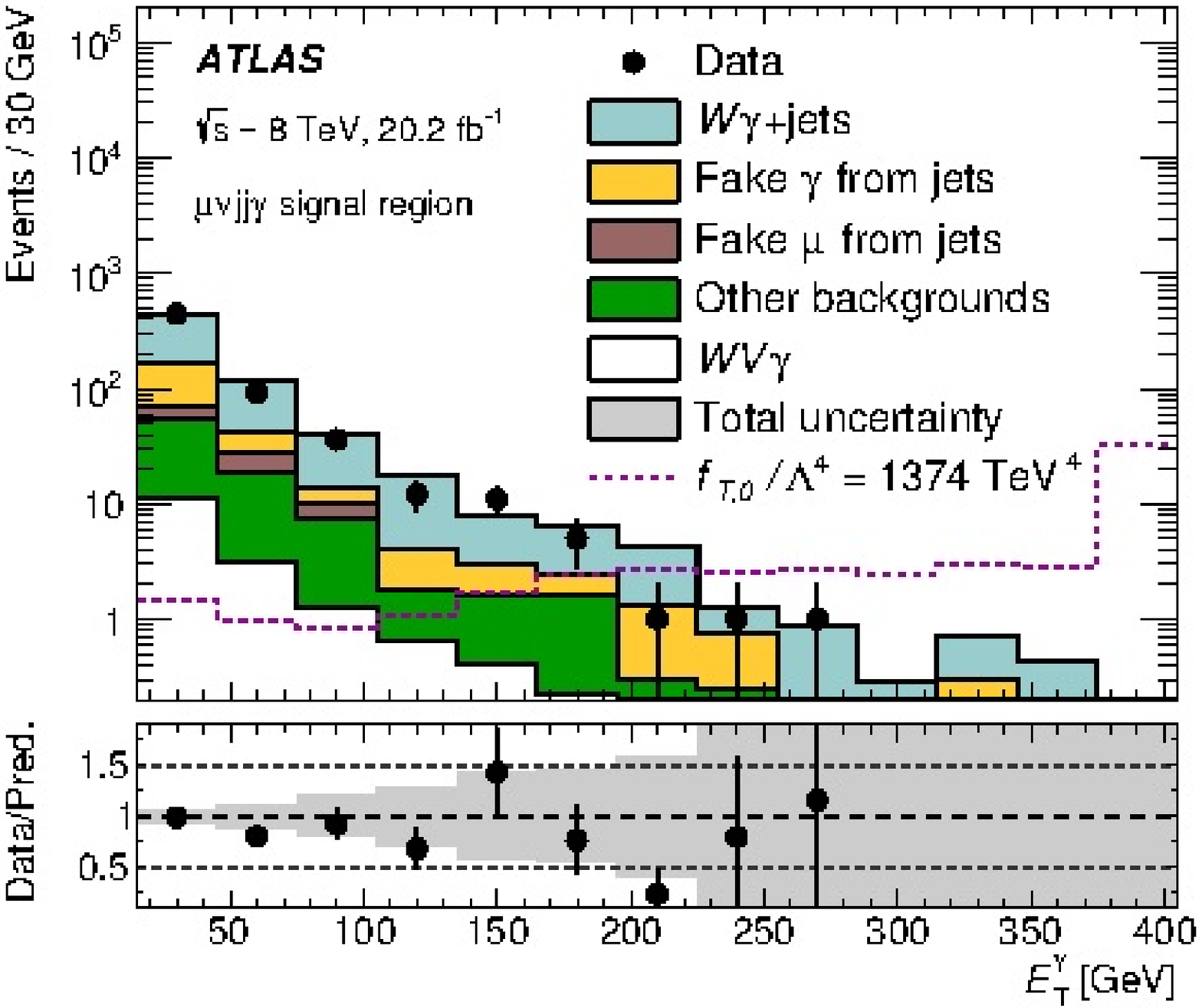}\label{fig:mjjy}}
 \caption{Observed and expected photon transverse energy distribution in the three signal regions. The data are shown together
with the signal and background expectations. Also indicated is the expected event yield for a reference model describing new 
physical phenomena (dashed histogram). The last bin contains all overflow events. The lower panel shows the ratio of the observed 
number of events and the sum of expected signal and background events as well as the uncertainties associated with the 
expectations. From Reference~\cite{Aaboud:2017tcq}.}
  \label{fig:results}
\end{figure}

\section{Results}
Figure~\ref{fig:results} shows the photon transverse energy distribution in the three signal regions and for all final states; 
the observed number of events is in good agreement with the expectation.
The production cross-section of the $e\nu\mu\nu\gamma$ final state is: 
$\sigma_{\mathrm{fid}}^{e\nu\mu\nu\gamma} = 1.5 \pm 0.9 (\mathrm{stat.}) \pm 0.5 (\mathrm{syst.})\,\mathrm{fb}$. 
It is in good agreement with the SM expectation of $\sigma_{\mathrm{theo}}^{e\nu\mu\nu\gamma} = 2.0 \pm 0.1$ computed at 
next-to-leading order using the VBFNLO program~\cite{Baglio:2014uba, Arnold:2011wj, Arnold:2008rz}.
Due to the lower significance, no cross-section is determined in the other two final states, but upper exclusion limits on their 
production cross-section are set using the CL$_{\mathrm{s}}$ method. Also the production cross-section of the $\ell\nu jj\gamma$ 
final state, being the combination of the $e\nu jj\gamma$ and $\mu\nu jj\gamma$ channels, is restricted and the limits at 95\% 
confidence level are as low as 2.5 times the cross-section expectation of the SM. 

The results are interpreted using an effective field theory that expands the Lagrangian density of the SM by operators 
of dimension eight. This introduces anomalous quartic gauge couplings leading to enhanced triboson production at high transverse 
photon energies as can be seen by the dashed histograms in Figure~\ref{fig:results}. 
However, unitarity can be violated in these descriptions and a dipole form factor is used in this interpretation to mitigate the 
effect~\cite{Degrande:2013rea}. 
To increase the sensitivity of this study, the photon \et\ threshold is increased to 120\,GeV for the $e\nu\mu\nu\gamma$ final 
state and to 200\,GeV for the $e\nu jj\gamma$ 
and $\mu\nu jj\gamma$ final states.
All final states are combined to constrain the coupling strength of 14 different dimension eight operators. 
These confidence intervals are computed for three different values of the scale $\Lambda_{FF}$ of the dipole form factor, namely 
$\infty$, 1 TeV, and 0.5 TeV. 
These results confirm and extend previous limits.
In addition, exclusion limits on the production cross-section for all final states and for the $\ell\nu jj\gamma$ final state 
with the raised photon \et\ threshold are computed again using the CL$_{\mathrm{s}}$ method. 

\section{Conclusions}
The $e\nu\mu\nu\gamma$, $e\nu jj\gamma$ and $\mu\nu jj\gamma$ final states of $WW\gamma$ and $WZ\gamma$ production are used to 
test the Standard Model; no deviations from the SM expectations are found. 
The production cross-section of the $e\nu\mu\nu\gamma$ final state is determined and the production cross-section of the $e\nu 
jj\gamma$ and $\mu\nu jj\gamma$ final states as well as their combination is constrained. 
In addition, confidence intervals for the coupling strength of 14 operators of dimension eight are derived using three different 
choices for the unitarisation.

\end{document}